\documentclass[doublecol,comment]{epl2} 
\usepackage[utf8]{inputenc}
\usepackage[english]{babel}
\usepackage{amsmath,amssymb,graphicx,enumerate,bm,physics}

\newcommand{\calB}{\mathcal{B}}

\title{Comment on ``Critique and correction of the currently accepted
  solution of the infinite spherical well in quantum mechanics'' by Huang
  Young-Sea and Thomann Hans-Rudolph}
\shorttitle{Comment on ``Critique and correction of the currently
  accepted solution of the infinite spherical well in QM'' } %Insert here a short version of the title if it exceeds 70 characters

\author{Antonio Prados\inst{1} \and Carlos A.~Plata\inst{1}}
\shortauthor{A.~Prados and C.~A.~Plata}

\institute{                    
  \inst{1} Física Teórica, Universidad de Sevilla, Apartado de Correos
  1065, E-41080 Sevilla, Spain
}
\begin{document}

\maketitle

%\section{Section title}

In a recent paper \cite{HyT16}, Huang and Thomann criticise the
``traditional'' solution of the time-independent Schr\"odinger
equation for a particle of mass $\mu$ in an infinite spherical well,
\begin{equation}\label{eq:schrod}
H\psi(\bm{r})\equiv\left[-\frac{\hbar^{2}}{2\mu}\nabla^{2}+V(r) \right]\psi(\bm{r})=E\psi(\bm{r}),
\end{equation}
where $V(r)=0$ for $r\leq a$, $V(r)=\infty$ for $r>a$. The wave
function, as usual, is written in spherical coordinates as
$\psi_{klm}(\bm{r})=R_{kl}(r)Y_{l}^{m}(\theta,\phi)$, where
$R_{kl}(r)$ and
$Y_{l}^{m}(\theta,\phi)$ are the radial part of the wave function and
the spherical harmonics, respectively. Then,
$\chi_{kl}(r)\equiv r R_{kl}(r)$ satisfies % the radial equation
\begin{equation}
  \label{eq:radial-eq}
  \frac{d^{2}\chi_{kl}}{dr^{2}}+\left[k^{2}-\frac{2\mu V(r)}{\hbar^{2}}-\frac{l(l+1)}{r^{2}}\right]\chi_{kl}=0,
\end{equation}
in which $k=\sqrt{2\mu E}/\hbar$. 

Their criticism may be summarised as follows: Since the particle is
confined to the spherical well, the region outside  is
irrelevant and the Schr\"odinger equation is solved inside the well,
that is, in the region
$\calB_{a}\equiv \left\{\bm{r}\in\mathbb{R}^{3}|0\leq r\leq
  a\right\}$ where $V(r)=0$.
Consistently, the derivatives of the wave function are treated by
one-sided limits at the boundaries. Equation \eqref{eq:radial-eq} has
two linearly independent solutions $rj_{l}(kr)$ and $rn_{l}(kr)$,
being $j_{l}$ and $n_{l}$  the spherical Bessel and Neumann
functions, respectively. Firstly, for $l>0$, $rn_{l}(kr)$ is not
square-integrable, and thus the only acceptable solution is
$j_{l}(kr)$. Secondly, for $l=0$, $rn_{0}(kr)$ cannot be disregarded,
as it is usually done, because $rn_{0}(kr)=-\cos(kr)$ is a
square-integrable function in the interval $0\leq r\leq a$. Thirdly,
any two solutions of \eqref{eq:radial-eq} must verify the
relation\footnote{Equation (15) in Ref.~\cite{HyT16}.}
\begin{equation}
  \label{eq:H-condition}
  \left.\chi_{kl}^{*}(r)\frac{d\chi_{k'l}(r)}{dr}- \frac{d\chi_{kl}^{*}(r)}{dr}\chi_{k'l}(r)\right|_{0}^{a}=0,
\end{equation}
which holds for any $(k,k',l)$ and stems from imposing that
$\matrixel{\psi_{klm}}{H}{\psi_{k'l'm'}}=\matrixel{\psi_{k'l'm'}}{H}{\psi_{klm}}^{*}$. % In \eqref{eq:H-condition},
% we have introduced the notation
% \begin{equation}
%   \label{eq:chi-def}
%   \chi_{kl}(r)=rR_{kl}(r).
% \end{equation}

Huang and Thomann show that \eqref{eq:H-condition} is compatible with
the eigenfunction $\chi_{k0}(r)=rn_{0}(kr)$ and calculate new energy
levels and stationary states corresponding to the ``conventional''
boundary condition $\chi_{kl}(a)=0$. Furthermore, they conclude by
stating that (i) at the edge of the wall, $\chi_{kl}(a)=0$ is not the
only possible boundary condition\footnote{Remember that the wave
  function is only defined inside $\calB_{a}$, and thus the usual
  continuity arguments at $r=a$ cannot be applied.} and (ii) thus a
major open problem is to find the correct boundary condition by sound
physical arguments.

In the following, we revisit this quantum mechanical problem and show
that the ``traditional'' solution is indeed the right one, although
some care is needed when deriving it if the wave function is
restricted to the region $\calB_{a}$, as it is done in \cite{HyT16}.  In
order to prove this, we make use of (i) the hermiticity of the
operator representing the radial component of the linear momentum,
that is,
\begin{equation}\label{eq:pr}
p_{r}=\frac{1}{2}\left(\bm{p}\cdot\frac{\bm{r}}{r}+
\frac{\bm{r}}{r}\cdot\bm{p}\right)=-i\hbar
\left(\frac{\partial}{\partial r}+\frac{1}{r}\right),
\end{equation}
and (ii) the fact that $\psi_{klm}(\bm{r})$ must be a solution of the
Schr\"odinger equation in the considered region, \textit{including the
  origin} $\bm{r}=\bm{0}\in \calB_{a}$.  Our proof follows the line of
reasoning in \cite{Messiah} to derive the boundary condition
$\chi_{kl}(0)=0$ for a particle moving in a central potential, but
with the (new) restriction of considering the wave function defined
only in the region $\calB_{a}$.

It is straightforward to show, integrating by parts in the interval
$0\leq r\leq a$, that
\begin{eqnarray}
  \matrixel{\psi_{klm}}{p_{r}}{\psi_{k'l'm'}}\!\!\!\!\!&=&\!\!\!\!\!
\matrixel{\psi_{k'l'm'}}{p_{r}}{\psi_{klm}}^{*} \nonumber \\
&& \!\!\!\! -\left.
i\hbar\,\chi_{kl}^{*}(r)\chi_{k'l}(r)\right|_{0}^{a}\delta_{ll'}\delta_{mm'}.
\label{eq:herm-pr}
\end{eqnarray}
% Note that we have $(klm)$ and $(k'lm)$, and not $(k'l'm')$, because
% $p_{r}$ skips the angular part $Y_{l}^{m}$ of the wave function and thus
% $\matrixel{\psi_{klm}}{p_{r}}{\psi_{k'l'm'}}=0$ if
% $(l,m)\neq (l',m')$. 
% Therefore, one has that
% \begin{equation}
%   \label{eq:herm-pr-conc-gen}
%   \chi_{kl}^{*}(0)\chi_{k'l}(0)=\chi_{kl}^{*}(a)\chi_{k'l}(a), \quad
%   \forall (k,k',l).
% \end{equation}
Particularisation of \eqref{eq:herm-pr} for $k=k'$, $l=l'$, $m=m'$ gives
\begin{equation}
  \label{eq:herm-pr-conc-diag}
  |\chi_{kl}(0)|=|\chi_{kl}(a)|, \quad
  \forall (k,l).
\end{equation}
This is the main conclusion stemming from the hermiticity of
$p_{r}$.

Note that, for $l>0$, \eqref{eq:herm-pr-conc-diag} makes it possible
to derive the only correct boundary conditions for $r=0$ and
$r=a$. Taking into account that the Neumann functions cannot be used
because they lead to non-normalisable wave functions,
$\chi_{kl}(r)=rj_{l}(kr)$. Then,
$\chi_{kl}(0)=\lim_{r\to 0}rj_{l}(kr)=0$, and then
\begin{equation}\label{eq:bc-r=a-l-non-zero}
  \chi_{kl}(0)=\chi_{kl}(a)=0, \quad \forall (k,l> 0).
\end{equation}

The case $l=0$ must be treated with greater care: we impose that the
wave function must be a solution of the Schr\"odinger equation in
$\calB_{a}$. We show below that this
excludes radial wave functions $R_{kl}(r)$  behaving as
$r^{-1}$ for $r\to 0$.  Solving \eqref{eq:radial-eq} gives
\begin{eqnarray}\label{eq:rad-eq-sol-r-small}
R_{k0}(r)&=& \frac{A \sin(kr)+B\cos(k
  r)}{r} \nonumber \\
&=&\frac{B}{r}+\frac{A \sin(kr)+B\left[\cos(k
  r)-1\right]}{r},
\end{eqnarray}
where $A$ and $B$ are arbitrary constants. The second term on the rhs
of \eqref{eq:rad-eq-sol-r-small} is regular at the origin, and no
problem stems from the application of the Laplacian operator
thereto. On the contrary, the first term is singular, in fact
$\nabla^{2}(1/r)=-4\pi\delta(\bm{r})$, where $\delta(\bm{r})$ is the
Dirac delta function. This implies that
\begin{equation}
  \label{eq:schrod-eq-r-small}
  \nabla^{2}\psi_{k00}(\bm{r})=
  -k^{2}\psi_{k00}(\bm{r})-B(4\pi)^{1/2}\delta(\bm{r}),
\end{equation}
where we have taken into account that
$\psi_{k00}(\bm{r})=R_{k0}(r)Y_{0}^{0}$, and
$Y_{0}^{0}=(4\pi)^{-1/2}$. Therefore, it is in order to have that
$\psi_{k00}(\bm{r})$ solves the Schr\"odinger equation in $\calB_{a}$,
\textit{including the origin}, that one must impose $B=0$. Note that
there is no delta contribution in the Schr\"odinger equation
\eqref{eq:schrod}. % Note that \eqref{eq:rad-eq-sol-r-small} and
% \eqref{eq:schrod-eq-r-small} are stronger for the spherical well,
% since they are exact equalities and not only asymptotic ones close to
% the origin.

The discussion in the above paragraph precludes $1/r$ terms in the
solutions of the Schr\"odinger equation in $\calB_{a}$. Thus, we
have that the only acceptable solution for the $l=0$ radial function
in the spherical well is
\begin{equation}\label{eq:schrod-sol-l=0}
  R_{k0}(r)=A\frac{\sin (kr)}{r},
\end{equation}
which implies that
% \begin{equation}\label{eq:conc-1}
$\chi_{k0}(0)=\lim_{r\to 0}rR_{k0}(r)=0.$
% \end{equation}
Once more, making use of \eqref{eq:herm-pr-conc-diag} one obtains
\begin{equation}\label{eq:conc-2}
\chi_{k0}(0)=\chi_{k0}(a)=0, \quad \forall k.
\end{equation}

Therefore, the main conclusion of our analysis for the
spherical well reads 
\begin{equation}\label{eq:general-conc}
  \chi_{kl}(0)=\chi_{kl}(a)=0, \quad \forall (k,l).
\end{equation}
This means that the only physically sound boundary conditions are the
``conventional'' ones. Moreover, \eqref{eq:H-condition} is always fulfilled. 
% Restricting the physical problem to a confined interval instead of the
% whole space could bring about some ambiguity in the boundary
% conditions. Nevertheless, our result \eqref{eq:general-conc} shows
% that it is not the case for the infinite spherical well. Indeed,

The above result can be extended to quite general 
spherically symmetric potentials. A key point is that
\eqref{eq:herm-pr-conc-diag} does not use the specific $r$-dependence
of $V(r)$. Let us assume that a particle is restricted to move inside
$\calB_{a}$ under a potential $V(r)$ that is bounded therein, except
possibly at $r=0$, where it may diverge as fast as $r^{-1}$, and
$r=a$. Therefore, for $l>0$, \eqref{eq:bc-r=a-l-non-zero} remains
valid because % $V(r)$
% does not diverge faster than $r^{-2}$ for small $r$:
the asymptotics for small $r$ of the wave function is dominated by the
``centrifugal'' contribution $l(l+1)/r^{2}$ and thus is equal to the
$V(r)=0$ case considered here. Having shown that $\chi_{kl}(0)=0$,
$\chi_{kl}(a)=0$ follows directly from
\eqref{eq:herm-pr-conc-diag}. Also, our conclusion regarding the
boundary conditions for $l=0$ still holds. Specifically, near the
origin, a Frobenius series expansion of the radial equation shows that
$\psi_{k00}(\bm{r})\sim (c_{1}+c_{2} r^{-1})Y_{0}^{0}$ and thus
$\nabla^{2}\psi_{k00}(\bm{r})\sim -(4\pi)^{1/2} c_{2}\delta(\bm{r})$
if $c_{2}\neq 0$. This enforces $c_{2}=0$ in order to make it possible
for $\psi_{k00}(\bm{r})$ to solve the Schr\"odinger equation. Then,
$\chi_{k0}(0)=\lim_{r\to 0}rc_{1}=0$ and, again,
\eqref{eq:herm-pr-conc-diag} yields $\chi_{k0}(a)=0$.

% \eqref{eq:schrod-eq-r-small} can be generalised to
% $(H-E)\psi(\bm{r},l=0)\sim \frac{2C}{\mu}\pi\hbar^{2}\delta(\bm{r})$
% in the vicinity of the origin \cite{Messiah}, which makes it
% impossible to find such a behaviour.

Restricting the physical problem to a finite region instead of the
whole space entails that boundary conditions have to be treated
carefully. Nevertheless, our result \eqref{eq:general-conc} shows that
the ``conventional'' boundary conditions are the correct ones not only
for the infinite spherical well but also for quite general
three-dimensional spherically symmetric potentials, diverging at most
as $r^{-1}$ for $r\to 0$.  Therein, the energy eigenvalues and
stationary states are the ``traditional'' ones found in textbooks
\cite{Messiah,Griffiths} and there is no room for novel solutions of
the Schr\"odinger equation.

% Insert here the text.
% See fig.~\ref{fig.1}, table~\ref{tab.1} and eq.~(\ref{eq.1}).
% See also~\cite{b.a,b.b}.
% \begin{equation}
% \label{eq.1}
% 0\neq1
% \end{equation}

% \begin{figure}
% \onefigure{epl-template.eps}
% \caption{Figure caption.}
% \label{fig.1}
% \end{figure}

% \begin{table}
% \caption{Table caption.}
% \label{tab.1}
% \begin{center}
% \begin{tabular}{lcr}
% first  & table & row\\
% second & table & row
% \end{tabular}
% \end{center}
% \end{table}

\acknowledgments We acknowledge the support of the Spanish Ministerio
de Econom\'{\i}a y Competitividad through grant FIS2014-53808-P.
Carlos A. Plata also acknowledges the support from the FPU Fellowship
Programme of the Spanish Mi\-nis\-te\-rio de Educaci\'on, Cultura y Deporte
through grant FPU14/00241.

\end{document}